\documentclass[11pt]{article}
\usepackage{color}

\usepackage{latexsym}  
\usepackage{amssymb}
\usepackage{graphicx}
\usepackage{amsmath}
\usepackage{mathrsfs}

\newcommand{\be}{\begin{equation}}
\newcommand{\ee}{\end{equation}}
\newcommand{\bea}{\begin{eqnarray}}
\newcommand{\eea}{\end{eqnarray}}
\newcommand{\bref}[1]{(\ref{#1})}

\newcommand{\pa}{\partial}

\begin{document}
\begin{titlepage}

\begin{center}
{\Large\bf  
Revisiting the Origin of the Universe and the Arrow of Time.}
\end{center}

\begin{center}

\vspace{0.1cm}

{\large Takeshi Fukuyama$^{a,}$%
\footnote{E-mail: fukuyama@rcnp.osaka-u.ac.jp}}

\vspace{0.2cm}

{\small \it ${}^a$Research Center for Nuclear Physics (RCNP),
Osaka University, \\Ibaraki, Osaka, 567-0047, Japan}


\end{center}

\begin{abstract}
We reconsider the old but yet unsolved problems, origin of the universe and the arrow of time. We show that only the closed universe is free from the singularity with the arrow of time symmetric with respect to the maximal size of the cosmic scale. The Wheeler-DeWitt equation is explicitly solved to obtain the local dynamical times. Corresponding to these local dynamical times, the thermodynamic arrow of time is proved to coincide with the arrows of dynamical time and of expanding  universe (cosmological time). The proof is explicitly shown in two-dimensional spacetime.

\end{abstract}
\end{titlepage}
\section{Introduction}
What is the origin of the universe ?  Can it be free from the singularity \cite{Hawking1,Hawking2} ?  What is time ? These are the long-standing problems. Vilenkin proposed the model that the universe appears from nothing \cite{Vilenkin, Vilenkin2} in the closed universe, which seems to be nicely fitted with the noboundary condition of the universe \cite{Hawking3} and free from the singularity.
After the birth of the universe it just leads to the inflation, which may solve the problem why the early universe is in low entropy state \cite{Carrol}.
These facts come from the Wheeler-Dewitt equation \cite{Dirac,Wheeler,DeWitt}, which shows that the universe does not include the extrinsic time.
Then, under the closed universe, it gives rise to the problems how to derive the intrinsic time and how to determine the arrow of time in the expanding and contracting phases.
The early universe can be interpretted only after we solve the above mentioned problems comprehensively and consistently. This is the purpose of this paper.  The outline of this paper is as follows.
Section 2 is devoted to the short review of the birth of the universe in Vilenkin's scenario. Then we derive the dynamical time from the Wheeler-DeWitt equation in Section 3. Thermodynamic arrow of time is discussed in Section 4. Section 5 is devoted to discussions.

\section{Creation of the Universe from Nothing}
This section is a review of the creation of the universe from nothing by Vilenkin \cite{Vilenkin, Vilenkin2}, which is necessary to understand the motivation of this paper and a key ingredient of this paper.
Let us start with the following system of gravity plus scalar matter,
\be
S=\int \sqrt{-g}d^4x\left[-\frac{1}{16\pi G}R-\frac{1}{2}\partial_\mu\phi\partial^\mu\phi-V(\phi)\right].
\label{action}
\ee
If we consider the closed Friedman-Robertson-Walker (FRW) universe, the invariant line element is
\be
ds^2=dt^2-a^2(t)\left(d\chi^2+\sin^2\chi(d\theta^2+\sin ^2\theta d\varphi^2)\right),
\label{metric1}
\ee
where physical distance $r=a\sin \chi$.  In this paper we adopt $\hbar=c=1$ units.
Lagrangin of \bref{action} in the above metric takes the form of
\be
\mathcal{L}=\frac{3\pi}{4G}(1-\dot{a}^2)a+\pi^2a^3\dot{\phi}^2-2\pi^2a^3V(\phi).
\label{lagrangian}
\ee
Here and hereafter dot indicates the derivative with respect $t$. The Wheeler-DeWit equation of this system is
\be 
\mathcal{H}\Psi=0,
\label{WD}
\ee
where
\be
\mathcal{H}=-\frac{G}{3\pi a}P_a^2+\frac{1}{4\pi^2a^3}P_\phi^2-\frac{3\pi}{4G}a\left(1-\frac{8\pi}{3}Ga^2V(\phi)\right).
\ee
$P_a$ and $P_\phi$ are the canonical conjugates of $a$ and $\phi$, respectively. Then \bref{WD} indicates
\be
\left(\frac{\partial^2}{\partial a^2}+\frac{p}{a}\frac{\partial}{\partial a}-\frac{1}{a^2}\frac{\partial^2}{\partial\tilde{\phi}^2}-U(a,\phi)\right)\Psi=0.
\ee
Here $p$ is an order one parameter relevant to the order of $a$ and $P_a$, $\tilde{\phi}^2\equiv 4\pi G\phi^2/3$, and
\bea
U(a,\phi)&=&\left(\frac{3\pi}{2G}\right)^2a^2\left(1-\frac{8\pi}{3}Ga^2V(\phi_0)\right)\nonumber\\
&=&\left(\frac{3\pi}{2G}\right)^2a^2(1-H^2a^2)
\label{closed}
\eea
with $H^2\equiv \frac{8\pi G}{3}|V(\phi_0)|$. $V(\phi$ is 
\be
V(\phi)=\rho_v-\frac{1}{2}m^2\phi^2
\label{V}
\ee
with $V'(\phi_0)=0$, having a spontaneous symmetry breaking with a wrong mass sign. 
The scale factor satisfies the Friedman equation, 
\be
\dot{a}^2-H^2a^2=-k, ~~(k=1,0,-1~\mbox{for closed, flat, and open universes, respectively})
\label{friedman}
\ee
with k=1 in this case.
\begin{figure}[h]
\begin{center}
\includegraphics[scale=1.0]{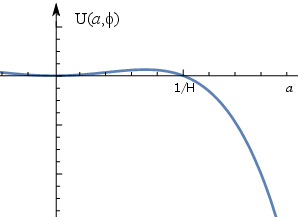}
\vspace{1.9cm}
\caption{
  The birth of the inflating universe from nothing \cite{Vilenkin}, which is possible only in the closed universe. The horizontal axis is the FRW scale factor $a$ and the vertical axis is $U(a,\phi)$  given in \bref{closed}. In the open universe, $U(a,\phi)$ is obtained by changing the signature of $H^2a^2$ in \bref{closed} and has no barrier to the singular point $a=0$. }
\label{fig:WD}
\end{center}
\end{figure}
This potential is depicted in Fig.1. It is very  important that, in the region $0<a<1/H$ (region I), $\dot{a}^2<0$ and the metric is Euclidean (time parameter $t$ is imaginary) and that, in the region $a>1/H$ (region II), the metric is Lorenzian.  

Before discussing furthermore on  this potential energy, let us make a comment on the implication of the closed universe. We can consider the case of open FRW universe simply by changing $a\to ia$ in Eq.\bref{closed} or $k=-1$ in Eq.\bref{friedman}. Its invariant length is
\be
ds^2=dt^2-a^2(t)\left(d\chi^2+\sinh^2\chi(d\theta^2+\sin ^2\theta d\varphi^2)\right).
\label{metric2}
\ee
In this case, the physical distance $r=a\sinh\chi$ increases without limit. As is easily seen from Eq.\bref{closed}, the whole region is Lorenzian and the universe shrinks into singularity in the early stage \cite{Hawking1, Hawking2}.

We proceed to discuss on the closed FRW universe in more detail. From \bref{friedman}, Lorenzian region is restricted in $a>1/H$ region. In region I the time is imaginary ($t=i\tau$), which means that $a=1/H$ region is connected with $a=0$ point by tunneling \cite{Vilenkin}. $a=0$ has no space-time and no energy, called ``nothing''. From \bref{friedman},
\be
a=H^{-1}\cos(H\tau).
\ee
After the tunneling in the real time, the scale factor $a$ behaves as
\be
a=H^{-1}\cosh Ht,
\label{inflation}
\ee
which implies the inflation. 
Thus the universe is free from the singularity and has no boundary \cite{Hawking3}. Due to the potential of \bref{V}, this inflation phase ends when $\phi$ reaches $\frac{m}{H}\phi\approx 1$ and the Big Bang begins. This also explains why the early universe is in low entropy state \cite{Carrol}. 

\section{The Dynamical Arrow of Time} 
We have reviewed that the closed universe is free from the singularity. However, it gives rise another problem whether $t=0$ is really the beginning of the universe over the whole history of it, leading to the question to which direction the arrow of time is directed in the contracting phase.

So far $t$ is really a parameter. \bref{WD} indicates that there is no time variable outside of the dynamical system, the universe.
We must introduce a $``time "$ variable intrinsic to the dynamical system.  
In this and next sections, we consider two different time arrows: dynamical time arrow and thermodynamic one. These two are mutually related as we will show.

In these two sections we consider the two dimensional gravity,  called Jackiw-Teitelboim (JT) gravity \cite{Jackiw,Teitelboim} which is amenable but reserves the universality of FRW metric in four dimension.
\be
S_{grav}=\int \sqrt{-g}(R-2\Lambda)N d^2x.
\label{2DG}
\ee
Here $R$ and $2\Lambda$ are scalar curvature and cosmological constant, respectively. It should be remarked that $R$ is total derivative in two dimensins. This action was formulated as $O(2,1)$ gauge theory by Fukuyama-Kamimura \cite{Kamimura1}. The usual four dimensional gravity is formulated as $O(4,1)$ or $O(3,2)$ gravity, which will be argued in Discussions.
In general, $g_{\mu\nu}$ is expressed as
\be
g_{\mu\nu}=e^{2\chi}
 \left(
  \begin{array}{cc}
   \eta_1^2-\eta_\bot^2&\eta_1\\
   \eta_1& 1\\
     \end{array}
 \right).
 \label{metric}
 \ee
 Then, $S_{grav}$ reads
 \be
 S_{grav}=\int (P_\chi\dot{\chi}+P_N\dot{N}-\eta_\bot\mathcal{H}_\bot-\eta_1\mathcal{H}_1) d^2x
 \ee
 with 
 \be
\label{WD2}
\mathcal{H}_\bot=P_NP_\chi+N'\chi'-N''-\Lambda Ne^{2\chi} ,
\ee
and diffeomorphism constraint,
\be
\mathcal{H}_1=P_\chi\chi'+P_NN'-P_\chi'.
\label{Diffeo}
\ee
Thus, $\eta_\bot$ and $\eta_1$ act as Lagrange multipliers and we have the Hamiltonian constraint (the Wheeler-DeWitt  equation) $\bref{WD2}\approx 0$ and the diffeomorphism constraint $\bref{Diffeo}\approx 0$.

We consider two dimensional space-time since it makes possible to solve the Wheeler-DeWitt equation explicitly and to induce the $``time "$ variable as will be shown \cite{Kamimura2}. 
We diagonalize $\mathcal{H}_\bot$ by using 
\be
\sigma_\pm\equiv \frac{\chi\pm N}{\sqrt{2}},
\ee
and their canonically conjugates
\be
\pi_\pm=\frac{P_\chi \pm P_N}{\sqrt{2}}
\ee
as
\be
\mathcal{H}_\bot=\frac{1}{2}(\pi_+^2-\pi_-^2)+\frac{1}{2}(\sigma_+'^2-\sqrt{2}\sigma_+'')-\frac{1}{2}(\sigma_-'^2-\sqrt{2}\sigma_-'')-\frac{\Lambda}{\sqrt{2}}(\sigma_+-\sigma_-)e^{\sqrt{2}(\sigma_++\sigma_-)}.
\ee
 Here $P_\chi$ and $P_N$ are variables canonically conjugate to $\chi$ and $N$, respectively. Dot (dash) implies the derivative with respect to $t~(x)$.
 However, if we use the FWR metric 
\be
ds^2=dt^2-a^2(t)dx^2,
\ee
then the metrics are reduced to
\be
\eta_\bot=e^{-\chi}\equiv \frac{1}{a},~~\eta_1=0.
\label{metric2}
\ee
Here
\be
\chi=\chi(t),~~N=N(t),
\ee
being independent on $x$, and $\bref{Diffeo}\approx 0$ is automatically satisfied. Equations of motion are
\bea
\label{EQM1}
&&\ddot{\chi}+\dot{\chi}=\Lambda,\\
&&\ddot{N}+\dot{\chi}\dot{N}-2\Lambda N=0.
\label{EQM2}
\eea
When 
\be
\Lambda=-\alpha^2<0,
\ee
with the initial condition,
\be
a(t=0)=0,
\ee
\bref{EQM1} gives the closed universe solution
\be
a=a_{max}\sin\alpha t,
\ee
and \bref{EQM2} does
\be
N=A\cos \alpha t.
\ee
Here $A$ is an intgral constant which is assumed to be positive. It should be remarked that $N$, overall factor of Lagrangian, changes its signature at the maximum radius ($a_{max}$) and that a variable with negative (positive) metric is transformed to that with positive (negative) metric. Corresponding to it, the dynamical time variable is induced locally in phases and not globally.
Amenable point of two dimensional gravity is that the Wheeler-DeWitt equation in the FRW metric,
\be
\frac{\pa^2\psi}{\pa N\pa \chi}+\Lambda Ne^{2\chi}\psi=0,
\ee
can be solved explicitly as
\be
\psi=c_0\mbox{exp}\left(\frac{c_1}{2}N^2+\frac{\alpha^2}{2c_1}e^{2\chi}\right),
\label{sol1}
\ee
where $\Lambda=-\alpha^2$. $c_0$ and $c_i$ are complex numbers. 

Then, we have the current conservation
\be
\pa_-j^-+\pa_+j^+=0.
\ee
Here
\be
j^-\equiv i(\psi^*\pa_-\psi-\psi\pa_-\psi^*),~~j^+\equiv -i(\psi^*\pa_+-\psi\pa_+\psi^*),
\ee
and 
\be
\pa_-\equiv \frac{\pa}{\pa\sigma_-}~~~\mbox{and}~~~\pa_+\equiv \frac{\pa}{\pa \sigma_+}.
\ee
Writing $c_1\equiv -\beta+i\gamma$, \bref{sol1} becomes
\be
\psi=c_0\mbox{exp}\left[\frac{i\gamma}{2}\left(N^2-\frac{\alpha^2}{|c_1|^2}e{2\chi}\right)-\frac{\beta}{2}\left(N^2+\frac{\alpha^2}{|c_1|^2}e^{2\chi}\right)\right].
\label{sol2}
\ee
This equation implies that the width of the wave packet is 
\be
\Delta N\approx \frac{1}{\sqrt{\beta}}
\label{sp1}
\ee
in the early universe and 
\be
\Delta e^\chi\approx \frac{|c_1|}{\alpha\sqrt{\beta}}
\label{sp2}
\ee
around $a_{max}$.  \bref{sp1} and \bref{sp2} indicate that the space-time relation,
\be
\sigma_+-\sigma_-\approx \frac{1}{\sqrt{\beta}}
\ee
and
\be
\sigma_++\sigma_-\approx \mbox{ln}\left(\frac{|c_1|}{\alpha\sqrt{\beta}}\right),
\ee
respectively. These asymptotiuc behaviours are depicted in Fig.2.
\begin{figure}[h]
\begin{center}
\includegraphics[scale=0.4]{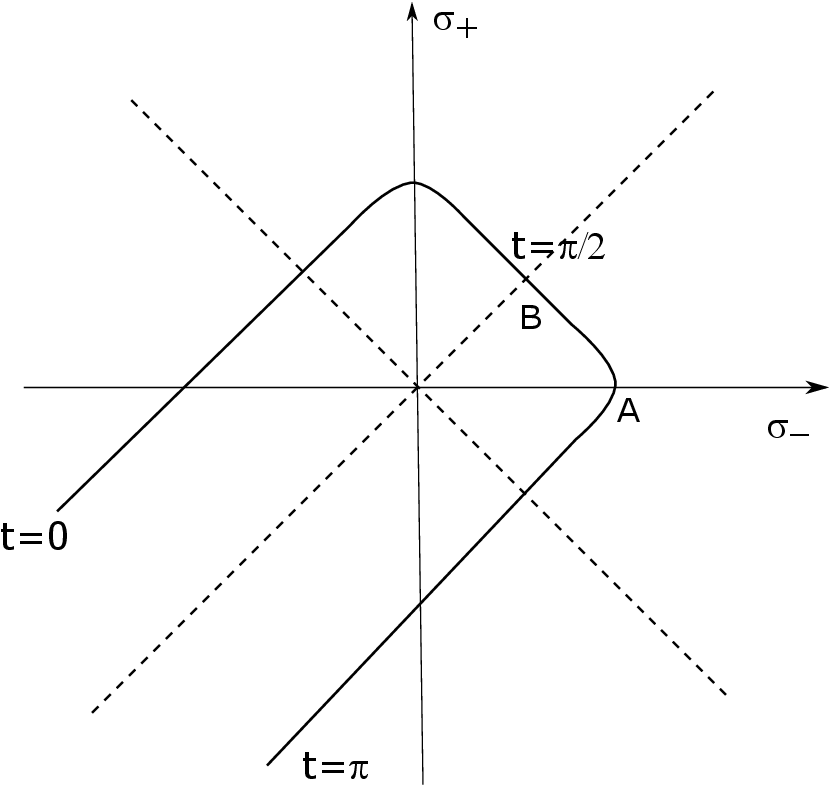}

\caption{ Classical trajectory in the $\sigma_--\sigma_+$ \cite{Kamimura2}. $B$ is the point at the maximal scale $a_{max}$.
  At the point $A$ and further, the condition $\dot{\sigma}_->0$ is broken.}
\label{fig:DT}
\end{center}
\end{figure}
In the semi-classical limit,
\be
\psi=ce^{iS},
\ee
where $S$ is the classical action \bref{2DG} in FRW metric. In this case, $j^-$ takes the form
\be
j^-=-2|c|^2\pa_-S=2|c|^2e^\chi\dot{\sigma}_-.
\label{sigmam}
\ee
Positive definiteness of the probability density is thus equivalent to that of $\dot{\sigma}_-(=-\pa_-S)$.
In the contracting phase, we should replace $\pa_-$ by $\pa_+$ in $j^-$,
\be
j^{(-)}\equiv j^-(\pa_-\to \pa_+)=-2|c|^2\pa_+S=-2|c|^2e^\chi\dot{\sigma}_+.
\label{sigmap}
\ee
Taking the behaviour of Fig.2 into consideration, the positive definiteness of $j^{(-)}$ is satisfied in the region from $t=0$ point to the point $A$. Over the point $A$, $\dot{\sigma}_-<0$, then ``time'' is replaced by $\sigma_+$ there.

\section{Thermodynamic Arrow of Time}
We have considered the thermodynamic arrow of time \cite{Morikawa} in the same system as \bref{2DG}.
We consider that the universe was under quantum theory in the vicinity of birth, whereas at present there seems to be no quantum correlation which, if it existed, drastically destroyed deterministic interpretation of cosmic observation. There must be a transition from an era which is fully quantum mechanical to an era which has no quantum coherence in the course of cosmological evolution.
We call this direction from the quantum to the classical era the thermodynamic arrow of time. More concretely speaking, this direction is that of the decrease of the dispersion (the width of the quantum coherence) $\sigma^2$ of the density matrix 
\be
\overline{\rho}[a_+,\eta_{\bot +},\phi_+;a_-,\eta_{\bot -},\phi_-]\equiv \psi(a_+,\eta_{\bot _+},\phi_+)\psi^*(a_-,\eta_{\bot -},\phi_-) .
\ee
Here $g_{\mu\nu}$ of \bref{metric} is parameterized as
\be 
g_{\mu\nu}=a^2\mbox{diag}(-\eta_{\bot}^2, 1)
\ee
and $\phi$ ia a scalar matter. Based on the influence functional method \cite{Feynman, Caldeira, Chou},
the reduced density matrix is given by
\be
\rho[a_+,\eta_{\bot +};a_-,\eta_{\bot -}]\equiv\int d\phi^+\int d\phi^-\overline{\rho}[a_+,\eta_{\bot +},\phi_+;a_-,\eta_{\bot -},\phi_-]\delta(\phi_+-\phi_-).
\ee
Then, the dispersion $\sigma$ is given by
\be
\rho[a_+;a_-]\propto exp[-\frac{(a_+-a_-)^2}{2\sigma^2}].
\label{dispersion}
\ee
The detailed calculations are given in \cite{Morikawa}.
The dispersion versus the scale factor $a$ is depicted in Fig.3. Thus the dispersion is decreasing in the expanding phase but
turns to increase at the contracting phase. Since the thermodynamic arrow of time is directed to that of the decreasing dispersion, it means that the arrow of the time changes its direction at $a_{max}$ (that is, the arrow of time is symmetric with respect to $a_{max}$).  It appears to decrease again beyond the point $A$ as the parameter $t$ increases. However, it is fictitious since in this region $\dot{\sigma}_-<0$ and the dynamical time should be replaced by $\sigma_+$ instead of $\sigma_-$ as we explained at the end of the previous section.  It should be remarked that in the contracting phase ($\frac{\pi}{2}<t<\pi$)
\be
\sigma_+(t)=\sigma_-(\pi -t),~~\sigma_-(t)=\sigma_+(\pi -t),
\ee
which implies that the direction of the dynamical time changes the signatute. Then the gradient of the dispersion changes its sign correspondingly and the dispersion increases in the region $AO$ as the time $\sigma^+$ increases.

\begin{figure}[h]
\begin{center}
\includegraphics[scale=0.4]{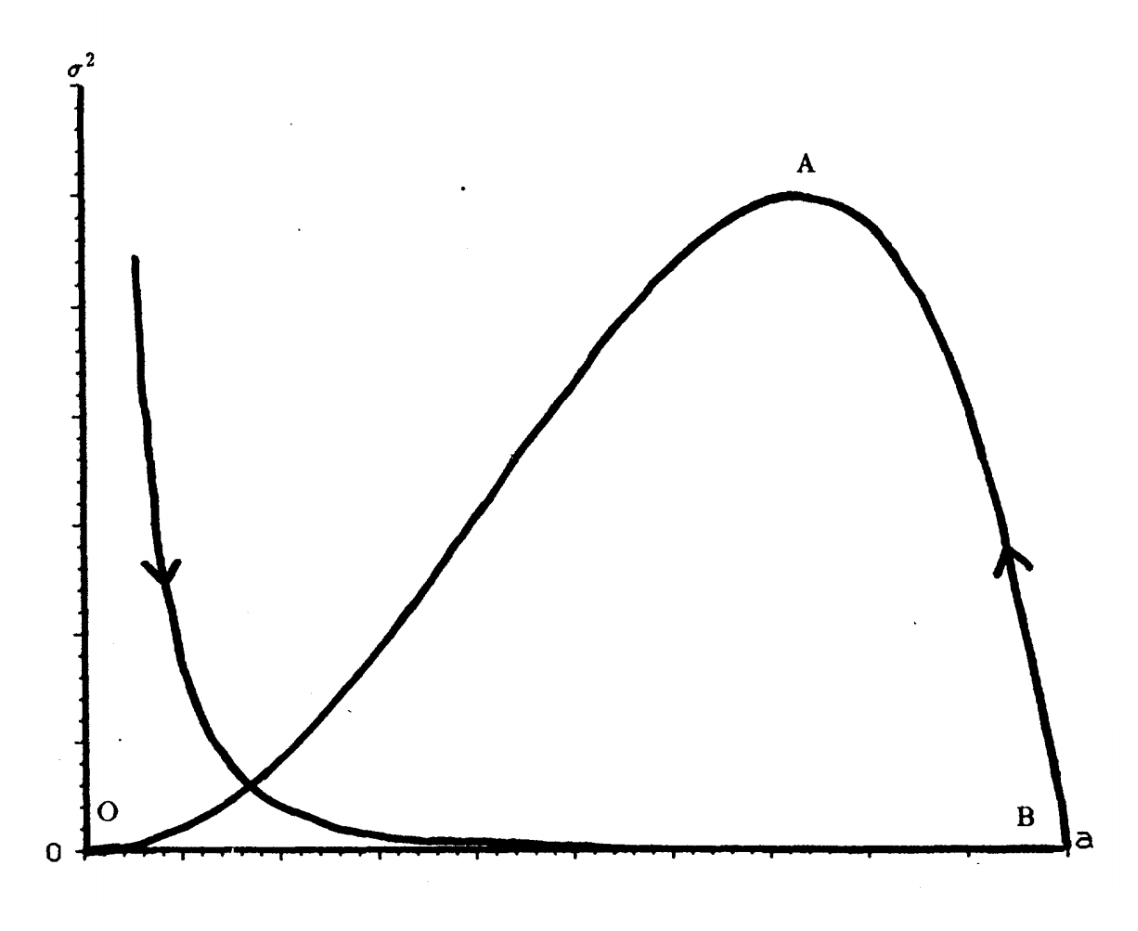}

\caption{Quantum coherence width $\sigma^2$ versus the scale factor $a$ is shown for the closed universe \cite{Morikawa}. $A$ and $B$ points correspond to those of Fig.2. }
\label{fig:Decoherence}
\end{center}
\end{figure}

On the contrary, Hawking et al. asserted that the universe does not change  its arrow of thermodynamic time in the contracting phase, asymmetric with respect to the maximal radius, $a_{max}$ \cite{Laflamme}.
In their work, they explicily showed that one of the dynamical freedoms, massive scalar amplitude $a_n$, continues to increases over the $a_{max}$. However, it is not clear whether it leads to time symmetry as a whole system.  On the other hand, we can solve the Wheeler-DeWitt equation explicitly and can define the intrinsic time variable in Sec.3, and obtained the thermodynamic entropy in this section.
 It should be noted in our case the dispersion $\sigma^2$ is decreasing again in $AO$ region in Fig.3, seemingly leading to the same result as that of \cite{Laflamme}, except for the region $BA$. However, as we have mentioned above, it is fictitious. If we correctly replace the time variable from $\sigma_-$ to $\sigma_+$ in the region $AO$, the dispersion $\sigma^2$ continues to increase as in the region of $BA$. The global use of the $``time "$, $\sigma_-$, in the whole region leads to the negative probability in the region $AO$ as was discussed in \bref{sigmam}, \bref{sigmap}.
Also, if the time was asymmetric with respect to $a_{max}$, the universe is bounced at $a=1/H$ (since the tunneling is rare), and gave rise to another problem why we are not in one of such bouncing phases and why our universe was in low entropy state in the early universe.

\section{Discussions}
We have discussed that the dynamical time variables are defind locally (Sec.3), which is consistent with the thermodynamic arrow of time symmetric around $a_{max}$ (Sec.4).  These arguments are due to two dimensions, which allows the explicit calculation without approximations.  The arguments of Sec. 2 are those in mini-superspace, effectively also in two dimensions. We needs some arguments depending on four dimensional space-time.

In four dimensions, JT gravity of \bref{2DG} is generalised to \cite{MM, Fuku2}.
\be
\mathcal{L}_{gravity}=\pm \epsilon^{abcd}\epsilon^{\mu\nu\rho\sigma}\hat{R}_{\mu\nu ab}\hat{R}_{\rho\sigma cd}/(16g^2).
\label{4DG}
\ee
Here Latin and Greek letters indicate world and local Lorentz coordinates, respectively. $\epsilon^{abcd}$ is fully antisymmetric tensor, and
\be
\label{Rhat}
\hat{R}_{\mu\nu ab}=R_{\mu\nu ab}+e_{[\mu a}e_{\nu] b}/l^2,
\ee
where
\be
R_{\mu\nu ab}=\pa_{[\mu}\omega_{\nu]ab}-\omega_{[\mu a}{}^c\omega_{\nu]cb}.
\label{R}
\ee
Here $\omega_{\mu ab}$ and $e_{\mu a}$ are the spin connection and tetrad, respectively, and $e_{[\mu a}e_{\nu]b}\equiv e_{\mu a}e_{\nu b}-e_{\nu a}e_{\mu b}$. $g$ and $l$ are the gauge coupling constant and the length scale characterizing (anti-) de-Sitter gauge group, respectively. The detail was shown in \cite{Fuku2}.
Substituting \bref{Rhat} and \bref{R} into \bref{4DG}, we obtain
\be
\mathcal{L}_{gravity}=\pa_\mu K^\mu-\frac{1}{16\pi G}e\left(R+\frac{6}{l^2}\right),
\ee
where
\be
e=\mbox{det}e_{\mu a}=\sqrt{-g},~R=e^{\mu a}e^{\nu b}R_{\mu\nu ab}, ~~16\pi G=g^2 l^2.
\ee
The first quadratic term of $R_{\mu\nu ab}$ in \bref{4DG} is the topological invariant and a total derivative.
Thus, gravity is formulated as the quadratic field-strength like the other conformal gauge theories but dynamically survive the linear Einstein's action and cosmological constant due to the topological invariant.
The scale parameter $l$ or equivalently the cosmological constant $l^2/6$ comes from the breaking of the conformal gauge symmetry $O(4,2)$ \cite{Fuku2}. Thus, this Lagrangian leads us to the model of Sec.2 in the FRW metric \bref{lagrangian} with the effective cosmological constant due to the homogeneous Higgs potential and a strict constant $\Lambda$.
This cosmological constant makes inflate the universe and becomes the reason why the universe was born in the low entropy state \cite{Carrol}. Here the cosmological constant is exactly ``constant " as the breaking scale of the $O(4,2)$ symmetry. Anyhow, we have not shown explicitly in this paper that the arguments of sections 3 and 4 is valid in four dimensional space-time. However, it is suggestive that the spectral dimension of the quantum universe as function of the diffusion
time moves from two to four continuously under the definite arrow of time \cite{Loll}.


Lastly we have three comments:
The first comment is on the so-called Aharanov-Bohm time variable \cite{AB},
\be
T=\frac{1}{2}M\left(a\frac{1}{P_a}+\frac{1}{P_a}a\right),
\label{AB}
\ee
which is the canonically conjugate to the Hamiltonian,
\be
\mathcal{H}=\frac{P_a^2}{2M}
\ee
with the total mass of the universe, $M=\frac{3\pi a_{max}}{2G}$.
This ``time'' variable has been discussed from the motivation different from that of this paper and leading to the weak measurement \cite{AB2}, which contradicts with the usual Copenhagen interpretation of quantum mechanics \cite{Landau}.
Anyhow it is suggestive that this ``time'' is not smoothly behaved at $P_a\approx 0$ at $a_{max}$, unlike the asymmetric time with respect to $a_{max}$.

The second comment is on the Conformal Cyclic Cosmology \cite{Penrose}. They claimed that the ${\mathscr I}^-$ of the next cyclic universe can be immersed in the past ${\mathscr I}^+$ by making a conformal rescaling the latter ${\mathscr I}^+$.
However, such conformal transformation is not the coordinate transformation and conformally transformed space-time is different from the original one. 

The last comment is on the relation between our quantum decoherence and AdS/CFT theory \cite{Maldacena}
The important ingredient of the AdS/CFT correspondence is that larger entanglement entropy in CFT corresponds to the nearer distance in AdS.
The entangle entropy in AdS/CFT corresponds to the quantum coherence in our theory \bref{dispersion}, and the larger quantum coherence leads to the smaller (negative direction of intrinsic) time in the same space-time. Important difference of our theory from AdS/CFT is that our theory has no extrinsic time variable.

\section*{Acknowledgements} 
We would like to thank K. Kamimura and M. Morikawa. This paper is greatly indebted to the works in collaboration with them.

\end{document}